\documentclass[prd,twocolumn,showpacs,floatfix,superscriptaddress]{revtex4}

\usepackage{graphicx}
\usepackage{epsfig}
\usepackage{bm}
\usepackage{amssymb}
\usepackage{float}
\usepackage{amsmath}
\usepackage{dcolumn}
\usepackage{cancel}
\usepackage[colorlinks]{hyperref}
\usepackage[usenames,dvipsnames]{color}
\hypersetup{
     breaklinks=true,
    pdfstartview={FitH},    
    colorlinks=true,       
    linkcolor=blue,          
    citecolor=red,        
    filecolor=magenta,      
    urlcolor=blue,           
    anchorcolor=green,      
    linktocpage=true
}

\def\doi{http://doi.org}
\def\arxiv{http://arxiv.org/abs}

\begin{document}

\title{Barrow holographic dark energy}

\author{Emmanuel N. Saridakis}
\email{msaridak@phys.uoa.gr}
\affiliation{National Observatory of Athens, Lofos Nymfon, 11852 Athens, 
Greece}
\affiliation{Department of Physics, National Technical University of Athens, 
Zografou
Campus GR 157 73, Athens, Greece}
\affiliation{Department of Astronomy, School of Physical Sciences, 
University of Science and Technology of China, Hefei 230026, P.R. China}

\begin{abstract}  
We formulate Barrow holographic dark energy, by applying the usual 
holographic principle at a cosmological framework, but using the 
Barrow entropy  instead of the standard  Bekenstein-Hawking 
one. The former is an extended black-hole entropy that arises due to 
quantum-gravitational effects which deform the black-hole 
surface by giving it an intricate, fractal form.
We extract  a simple  differential equation for the evolution of the dark 
energy density parameter, which possesses standard holographic dark energy as a 
limiting sub-case,
and we show that the scenario can describe  the 
universe thermal history, with  the sequence of matter and dark energy eras. 
Additionally, the new Barrow exponent $\Delta$ significantly affects the dark 
energy equation of state, and according to its value it can lead it to lie in 
the quintessence regime, in the phantom regime, or experience the phantom-divide 
crossing during the evolution.

\end{abstract}


\maketitle

\section{Introduction}

Holographic dark energy is an interesting  alternative scenario for the 
quantitative description of dark energy, originating from  the holographic 
principle 
\cite{tHooft:1993dmi,Susskind:1994vu,Bousso:2002ju,Fischler:1998st,
Horava:2000tb}. Starting from the 
 connection between the largest length of a quantum field theory  with its 
  Ultraviolet cutoff   \cite{Cohen:1998zx}, one can result to 
a vacuum energy of holographic origin, which at cosmological scales   
form   dark energy \cite{Li:2004rb,Wang:2016och}.
Holographic dark energy   proves to lead to interesting cosmological 
behavior, both at its simple  
\cite{Li:2004rb,Wang:2016och,Horvat:2004vn,Huang:2004ai,Pavon:2005yx,Wang:2005jx,
Nojiri:2005pu,Kim:2005at,
Wang:2005ph, Setare:2006wh,Setare:2008pc,Setare:2008hm}, as well as at its 
extended versions
\cite{Gong:2004fq,Saridakis:2007cy,  
Setare:2007we,Cai:2007us,Setare:2008bb,Saridakis:2007ns,Saridakis:2007wx,Jamil:2009sq,
Gong:2009dc, 
Suwa:2009gm,BouhmadiLopez:2011xi,Chimento:2011pk,Malekjani:2012bw,
Chimento:2013qja,
Khurshudyan:2014axa,
Landim:2015hqa,Pasqua:2015bfz,
Jawad:2016tne,Pourhassan:2017cba,Nojiri:2017opc,Saridakis:2017rdo,
Saridakis:2018unr,Aditya:2019bbk,Nojiri:2019kkp,Geng:2019shx,Waheed:2020cxw}, 
and  it is in 
agreement  with   
observational data 
\cite{Zhang:2005hs,Li:2009bn,Feng:2007wn,Zhang:2009un,Lu:2009iv,
Micheletti:2009jy,DAgostino:2019wko,Sadri:2019qxt,Molavi:2019mlh}.

The important step in the application of   holograpic principle at   
cosmological framework is that the universe horizon (i.e. largest distance) 
entropy is proportional to its area, similarly to the Bekenstein-Hawking 
entropy of a black hole. However, very recently Barrow was inspired by the 
Covid-19 
virus illustrations and he showed that 
quantum-gravitational effects may introduce  intricate, fractal features on the 
black-hole structure. This complex structure leads to finite volume but with 
infinite (or finite) area,  and therefore to a deformed 
black-hole entropy expression \cite{Barrow:2020tzx}
\begin{equation}
\label{Barrsent}
S_B=  \left (\frac{A}{A_0} \right )^{1+\Delta/2}, 
\end{equation}
where $A$ is the standard horizon area  and $A_0$ the Planck area. The 
quantum-gravitational deformation is therefore quantified by the 
 new 
exponent $\Delta$, with $\Delta=0$ 
corresponding to the   standard 
Bekenstein-Hawking entropy (simplest horizon structure), and with  $\Delta=1$ 
corresponding to   the most intricate and fractal structure.  Notice that the 
above quantum-gravitationally corrected entropy is different than the usual 
``quantum-corrected'' entropy with logarithmic corrections 
\cite{Kaul:2000kf,Carlip:2000nv}, however
it resembles    Tsallis 
nonextensive entropy
\cite{Tsallis:1987eu,Wilk:1999dr,Tsallis:2012js}, nevertheless the 
involved foundations and physical principles 
are completely different. 
 Finally, note that the above  effective fractal behavior does not arise 
from specific   quantum gravity calculations, but from  general simple  
physical principles, which adds to its plausibility and hence it is valid 
  as a first approach on the subject \cite{Barrow:2020tzx}.

In the present manuscript we are interested in constructing holographic dark 
energy, but using the extended, Barrow relation for the horizon entropy, 
instead of the usual     Bekenstein-Hawking one. Barrow holographic dark energy 
  possesses  usual holographic dark energy as a limit in the  $\Delta=0$ 
case, however in general it is a new scenario with richer structure and 
cosmological behavior.

\section{Barrow holographic dark energy}

In this section we construct  the scenario of Barrow holographic dark energy.
While standard 
holographic dark energy  is given by the inequality $\rho_{DE} 
L^4\leq S$, where $L$ is the horizon length, and under the imposition $S\propto 
A\propto L^2 $ 
\cite{Wang:2016och}, the use of Barrow entropy (\ref{Barrsent}) will lead to
 \begin{equation}
\label{FRWTHDE}
\rho_{DE}={C} L^{\Delta-2},
\end{equation}
with ${C}$ a parameter with dimensions  $[L]^{-2-\Delta}$. In the case
where 
$\Delta=0$, as expected, the above expression provides the 
standard holographic dark energy  $\rho_{DE}=3c^2 M_p^2 L^{-2}$ (here $M_p$ 
is the Planck mass), where ${C}=3 
c^2 
M_p^2$ and with
$c^2$ the model parameter. However, in the case where the deformation effects 
quantified by $\Delta$ switch on, Barrow holographic dark energy will depart 
from the standard one, leading to different cosmological behavior.

We
consider a flat   Friedmann-Robertson-Walker (FRW) 
  metric  of the form
\begin{equation}
\label{FRWmetric}
ds^{2}=-dt^{2}+a^{2}(t)\delta_{ij}dx^{i}dx^{j}\,,
\end{equation}
where $a(t)$ is the scale factor. Concerning the   largest length  
$L$  which 
appears in the expression of any holographic dark energy, although there are 
many possible choices, the most common in the literature is to use 
  the   future event horizon \cite{Li:2004rb} 
\begin{equation}
\label{FRWfuturehor}
R_h\equiv a\int_t^\infty \frac{dt}{a}= a\int_a^\infty \frac{da}{Ha^2},
\end{equation}
 with $H\equiv \dot{a}/a$ the Hubble parameter. Hence, substituting $L$ in   
  (\ref{FRWTHDE}) with     $R_h$ we obtain the energy 
density of Barrow holographic dark energy, namely
 \begin{equation}
\label{FRWTHDE2}
\rho_{DE}={C} R_h^{\Delta-2}.
\end{equation}

We consider that the universe is filled with the usual matter 
perfect fluid, as well as with the above holographic dark energy. 
The two  Friedmann equations are then written as 
 \begin{eqnarray}
\label{Fr1bFRW}
3M_p^2 H^2& =& \ \rho_m + \rho_{DE}    \\
\label{Fr2bFRW}
-2 M_p^2\dot{H}& =& \rho_m +p_m+\rho_{DE}+p_{DE},
\end{eqnarray}
where $p_{DE}$ is the pressure of  Barrow holographic dark energy, and 
$\rho_m$, $p_m$ 
 the energy density and pressure of   matter,  respectively. Additionally,
 for the matter sector we consider the standard
conservation 
equation  
\begin{equation}\label{rhoconservFRW}
\dot{\rho}_m+3H(\rho_m+p_m)=0.
\end{equation}
 Finally,  we introduce the density parameters  
 \begin{eqnarray}
 && \Omega_m\equiv\frac{1}{3M_p^2H^2}\rho_m
 \label{OmmFRW}\\
 &&\Omega_{DE}\equiv\frac{1}{3M_p^2H^2}\rho_{DE}.
  \label{ODE}
 \end{eqnarray}

 Using the density parameters, expressions 
(\ref{FRWfuturehor}),(\ref{FRWTHDE2}) give
  \begin{equation}\label{integrrelation}
\int_x^\infty \frac{dx}{Ha}=\frac{1}{a}\left(\frac{{C}}{3M_p^2H^2\Omega_{DE}}
\right)^{\frac{1}{ 2-\Delta}},
\end{equation}
 with  $x\equiv \ln a$. Considering the matter to be dust ($p_m=0$), 
from (\ref{rhoconservFRW})  
we obtain  $\rho_m=\rho_{m0}/a^3$, with $\rho_{m0}$ the present 
matter 
energy density, namely at $a_0=1$ (in the following the subscript ``0'' denotes 
the    value of a quantity at present).  Thus, substituting into (\ref{OmmFRW}) 
leads to 
$\Omega_m=\Omega_{m0} H_0^2/(a^3 H^2)$, from which, using   the Friedmann 
equation   $\Omega_m+\Omega_{DE}=1$, 
we acquire
 \begin{equation}\label{Hrel2FRW}
\frac{1}{Ha}=\frac{\sqrt{a(1-\Omega_{DE})}}{H_0\sqrt{\Omega_{m0}}}.
\end{equation}

Inserting (\ref{Hrel2FRW}) into expression (\ref{integrrelation}) we get the 
useful relation
  \begin{equation}\label{integrrelation2FRW}
\int_x^\infty \frac{dx}{H_0\sqrt{\Omega_{m0}}}  
\sqrt{a(1-\Omega_{DE})}    =\frac{1}{a}\left(\frac{{C}}{
3M_p^2H^2\Omega_{DE}}
\right)^{\frac{1}{2-\Delta}}.
\end{equation}
Differentiating 
(\ref{integrrelation2FRW}) with respect to $x=\ln a$ we result to
  \begin{eqnarray}\label{Odediffeq}
&&
\!\!\!\!\!\!\!\!\!\!\!\!\!
\frac{\Omega_{DE}'}{\Omega_{DE}(1-\Omega_{DE})}=\Delta+1+
Q
(1-\Omega_{DE})^{\frac{\Delta}{2(\Delta-2) } } \nonumber\\
&&\ \ \ \ \ \ \ \ \ \ \ \ \ \  \ \ \ \ \ \ \ \ \ \  \  \ \ \   \ \ \ 
\cdot(\Omega_{DE})^{\frac{1}{2-\Delta } } 
e^{\frac{3\Delta}{2(\Delta-2)}x},
\end{eqnarray}
 with
   \begin{equation}\label{Qdef}
Q\equiv (2-\Delta)\left(\frac{{C}}{3M_p^2}\right)^{\frac{1}{\Delta-2}} 
\left(H_0\sqrt{\Omega_{m0}}\right)^{\frac{\Delta}{2-\Delta}}
\end{equation}
a dimensionless parameter and   where primes denote derivatives with respect to 
$x$.

The above differential equation   determines the evolution of Barrow holographic 
dark energy for dust matter  in a flat universe. In the case 
  $\Delta=0$  
it coincides with the  usual holographic dark energy,  
i.e. 
$\Omega_{DE}'|_{_{\Delta=0}}= 
\Omega_{DE}(1-\Omega_{DE})\left(1+2\sqrt{\frac{3M_p^2\Omega_{DE}}{{C}}}
\right)
$, which has an analytic solution in   implicit  form  \cite{Li:2004rb}. 
However, in the general case of Barrow exponent   
  $\Delta$, Eq. (\ref{Odediffeq})
presents an   $x$-dependence and it has to be elaborated numerically.

Using the above relations we can additionally calculate the  equation-of-state 
parameter for Barrow holographic dark energy $w_{DE}\equiv p_{DE}/\rho_{DE}$.
Differentiation of  (\ref{FRWTHDE2}) leads to
$\dot{\rho}_{DE}=(\Delta-2){C} R_h^{ \Delta-3} \dot{R}_h$, with $\dot{R}_h$ 
calculated using (\ref{FRWfuturehor}) as 
$\dot{R}_h=H  R_h-1$, and where according to (\ref{FRWTHDE2}) $R_h$ can be 
eliminated in terms of $\rho_{DE}$ 
as 
$ R_h=(\rho_{DE}/{C})^{1/(\Delta-2)}$. Inserting this into the dark energy 
conservation equation $\dot{\rho}_{DE}+3H\rho_{DE}(1+w_{DE})=0$ (which is a 
straightforward consequence of the matter conservation (\ref{rhoconservFRW})), 
we acquire 
\begin{eqnarray} 
&&
\!\!\!\!\!\!\!\!\!\!\!\!\!\!\!\!\!\!\!\!\! 
(\Delta-2){C} 
\left(\frac{\rho_{DE}}{{C}}\right)^{\frac{\Delta-3}{\Delta-2}}
 \left[H  
\left(\frac{\rho_{DE}}{{C}}\right)^{\frac{1}{\Delta-2}}-1\right]\nonumber\\
&&\ \ \ \ \ \  \ \  \ \ \ \ \  \ \ \ \ \ \ \ \ \
+3H\rho_{DE}
(1+w_{DE})=0.
\end{eqnarray}
Hence, inserting $H$ from (\ref{Hrel2FRW}), and using    (\ref{ODE}) we finally 
obtain
\begin{equation}\label{wDEFRW}
w_{DE}=-\frac{1\!+\!\Delta}{3}
-\frac{Q}{3}
(\Omega_{DE})^{\frac{1}{2-\Delta } } (1\!-\!\Omega_{DE})^{\frac{\Delta}{
2(\Delta-2) } }
e^{\frac{3\Delta}{2(2-\Delta)}x}.
\end{equation}
Therefore, the evolution of $w_{DE}$ in terms of $x=\ln a$ is known, as long as 
$\Omega_{DE}$ is known from   (\ref{Odediffeq}). Lastly, in the standard case 
of  
$\Delta=0$, expression
(\ref{wDEFRW})   gives 
$w_{DE}|_{_{\Delta=0}}=-\frac{1}{3}-\frac{2}{3}\sqrt{\frac{3M_p^2 
\Omega_{DE}}{{C}}}$, which is the usual holographic 
dark energy result
\cite{Wang:2016och}.

\section{Cosmological evolution}

\begin{figure}[ht]
  \hspace{-1.cm}
\includegraphics[scale=0.46]{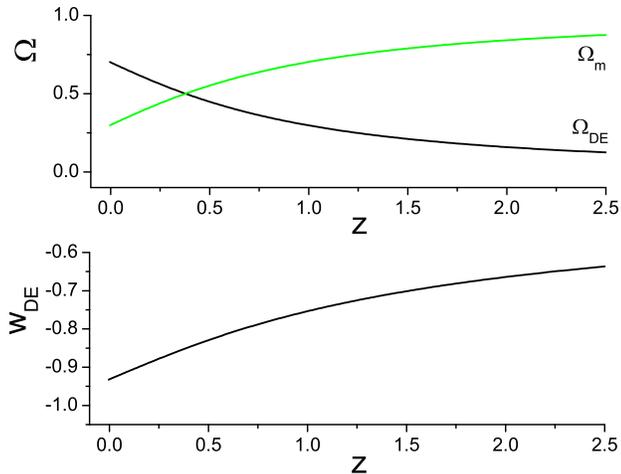}
\caption{
{\it{ Upper graph: The evolution of   matter and of  Barrow 
holographic dark energy density parameters, as a function of the redshift 
$z$, for $\Delta=0.2$ and ${C}=3$, in units where $M_p^2=1$. 
  Lower graph: The evolution of the corresponding dark-energy equation-of-state 
parameter $w_{DE}$.   We have imposed 
 $\Omega_{DE}(x=-\ln(1+z)=0)\equiv\Omega_{DE0}\approx0.7$  at present.
}} }
\label{OmegaFRWs}
\end{figure}

In this section we  investigate in detail the cosmological evolution in  the 
scenario of Barrow holographic dark energy. The dark energy density parameter 
$\Omega_{DE}$ is determined by Eq. (\ref{Odediffeq}), which can be solved 
analytically only in the standard case  $\Delta=0$ \cite{Li:2004rb}. 
Nevertheless, we can extract its solution through 
  numerical elaboration, and then find the    redshift behavior  knowing that 
$x\equiv\ln a=-\ln(1+z)$ (with $a_0=1$). 
Finally, concerning the initial conditions we impose  
$\Omega_m(x=-\ln(1+z)=0)\equiv\Omega_{m0}\approx0.3$ and thus 
$\Omega_{DE}(x=-\ln(1+z)=0)\equiv\Omega_{DE0}\approx0.7$ in agreement with 
observations 
\cite{Ade:2015xua}. In  Fig. \ref{OmegaFRWs} we depict the evolution of 
$\Omega_{DE}(z)$ and $\Omega_{m}(z)=1-\Omega_{DE}(z)$, as well as the 
corresponding evolution of $w_{DE}(z)$  from  
(\ref{wDEFRW}).  
As we observe the scenario at hand can successfully describe the 
thermal 
history of the universe, with  the sequence of matter and dark energy epochs. 
Moreover,  the value of $w_{DE}$ at present is around $-1$ as required 
  by observations.

\begin{figure}[ht]
 \hspace{-1.cm}
\includegraphics[scale=0.46]{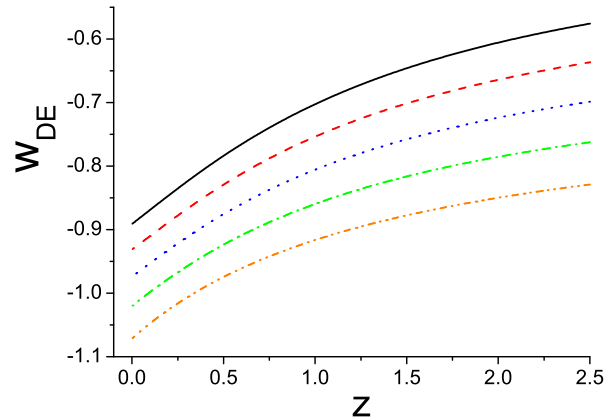}
\caption{
{\it{The   equation-of-state parameter $w_{DE}$ of Barrow
holographic dark energy, as a function of the redshift $z$, for ${C}=3$, and  
for 
 $\Delta=0$ (black-solid), $\Delta=0.2 $ 
(red-dashed), $\Delta=0.4$ 
(blue-dotted),    $\Delta=0.6$ (green-dashed-dotted), and $\Delta=0.8$ 
(orange-dashed-dot-dotted), in units 
where $M_p^2=1$.
We have imposed 
$\Omega_{DE}(x=-\ln(1+z)=0)\equiv\Omega_{DE0}\approx0.7$ at present.}} }
\label{wzplotFRW}
\end{figure}

Let us now investigate in more detail the equation-of-state parameter of 
Barrow holographic dark energy, and in particular to examine how it is affected 
by the Barrow exponent $\Delta$ that quantifies the deviation from the usual 
scenario. In Fig. \ref{wzplotFRW} we 
depict $w_{DE}(z)$ for various values of $\Delta$, including the standard value 
$\Delta=0$. A general observation is that for 
  increasing 
$\Delta$ 
the whole evolution of $w_{DE}(z)$, as well as its current value 
$w_{DE}(z=0)\equiv w_{DE0}$, tend to acquire lower 
values. We mention that for $\Delta \gtrsim  0.5 $ the value of 
$w_{DE0}$ lies in the 
phantom regime. This was expected, since 
expression (\ref{wDEFRW}) allows phantom values,
 which is a theoretical advantage of the scenario at hand and reveals 
its capabilities. Thus, as we see, according to the value of $\Delta$,  
Barrow holographic dark energy can lie in the quintessence or in the phantom 
regime,   or exhibit  the phantom-divide crossing 
during the cosmological evolution. 

We close this section by mentioning that the scenario of Barrow holographic 
dark energy has two parameters, i.e. the new Barrow exponent 
  $\Delta$, and the constant ${C}$ (similar to the parameter $c^2$ of standard 
holographic dark energy) which incorporates the initial inequality validation. 
In the above analysis we preferred to   fix  ${C}=3$, which is the value
required if we desire the present scenario to have standard 
holographic dark energy as an exact limit for  $\Delta=0$, and we examined the 
pure role of   $\Delta$ on the cosmological evolution. This was proved to be 
adequate for a successful description     in agreement with 
observations, which serves as a significant advantage comparing to standard 
holographic dark energy, in which case  one needs to adjust the value of the 
constant $c^2$   to  fit the data. Definitely, varying  the 
value of ${C}$ too would lead to even more improved cosmological behavior, 
which reveals  the capabilities of the scenario.

\section{Conclusions}
\label{Conclusions}

We constructed Barrow holographic dark energy, by applying the usual 
holographic principle at a cosmological framework,   but using the 
 Barrow entropy, instead of the standard  Bekenstein-Hawking 
one. Specifically, in a recent work Barrow proposed that quantum-gravitational 
effects may bring about intricate, fractal structure on the
black-hole surface, and hence lead to a deformed  black-hole entropy, 
quantified by a new exponent $\Delta$  \cite{Barrow:2020tzx}. Hence, the 
resulting Barrow holographic dark energy will possess the usual one as a limit, 
namely when  $\Delta=0$ which corresponds to the case where Barrow entropy 
becomes the standard one, but for  $\Delta>0$ and up to the maximal deformation 
for $\Delta=1$ it gives rise to novel cosmological scenarios.

We extracted  a simple  differential equation for the evolution of the dark 
energy density parameter, and we presented the solution for the evolution of 
the corresponding dark energy equation-of-state parameter. As we showed, the 
scenario of Barrow holographic dark energy can describe  the universe thermal 
history, with  the sequence of matter and dark energy eras. Additionally, the 
new Barrow exponent $\Delta$ significantly affects the dark energy equation of 
state, and according to its value it can lead it to lie in the quintessence 
regime, in the phantom regime, or experience the phantom-divide crossing during 
the evolution. The above behaviors were obtained by changing only the value of 
$\Delta$. Additional adjusting of the parameter ${C}$ will enhance
significantly the capabilities of the scenario.

 We would like to mention here that Barrow  entropy proposal is just a 
first 
approximation on the subject of quantum gravitational implications on the black 
hole horizons. In reality the underlying spacetime foam deformation will be 
complex, wild and dynamical. Nevertheless, as a first step  the complexity of 
the phenomenon can effectively and coarse-grained be embedded in the new 
exponent, and thus the highly dynamical deformation of the black hole 
surface can effectively  be described by $\Delta$, which is not fixed but it 
remains in an interval between extreme values. However, as a more realistic 
scenario which could incorporate the dynamicallity of spacetime 
foam, one could think of an exponent $\Delta$ that depends on time and scale, 
as it has already been done with Tsallis entropy exponent 
\cite{Nojiri:2019itp}.

Barrow holographic dark energy exhibits more interesting and richer 
phenomenology comparing to the standard scenario, and thus it can be a 
candidate for the description of nature. It would be both necessary and 
interesting to perform a full observational analysis, confronting the scenario 
with observational data from  Supernovae type Ia    (SNIa), Baryonic Acoustic 
Oscillations (BAO), and Cosmic Microwave Background (CMB) probes, as well as 
with  Large Scale Structure (such as f$\sigma$8)  data,  in order to  constrain 
the new parameter $\Delta$.   These necessary studies lie beyond the 
scope of the present work and are 
left for future investigation.

\end{document}